\documentclass[reprint,amsmath,amssymb,aps,prl]{revtex4-2}
\usepackage{amsmath}%
\usepackage{amsfonts}%
\usepackage{amssymb}%
\usepackage[cal=cm]{mathalfa}
\usepackage{graphicx}
\usepackage{dcolumn}
\usepackage{bm}
\usepackage{braket}
\usepackage{sidecap}
\usepackage{color}
\usepackage{bbm}
\usepackage{nicefrac}
\usepackage{float}
\usepackage{placeins}
\usepackage[dvipsnames]{xcolor}
\usepackage[utf8]{inputenc}
\usepackage[normalem]{ulem}
\usepackage{nicefrac}
\usepackage{dsfont}
\usepackage{xcolor, soul} 
\usepackage[normalem]{ulem} 
\usepackage{siunitx}
\usepackage[citebordercolor=blue]{hyperref}
\usepackage[utf8]{inputenc}
\usepackage{gensymb}
\usepackage{mathtools}

\newcommand{\myred}[1]{{\color{black} #1}}

\begin{document}

\title{Measuring the Angular Momentum of a Neutron Using Earth's Rotation}

\author{Niels Geerits}\email{niels.geerits@tuwien.ac.at}\author{Stephan Sponar}\affiliation{%
Atominstitut, Technische Universit\"at Wien, Stadionallee 2, 1020 Vienna, Austria}
\author{Kyle E. Steffen} \author{William M. Snow}\affiliation{Centre for Exploration of Energy and Matter, Indiana University, Bloomington,47408, US}
\author{Steven R. Parnell} \author{Giacomo Mauri} \author{Gregory N. Smith} \author{Robert M. Dalgliesh}\affiliation{ISIS, Rutherford Appleton Laboratory, Chilton, Oxfordshire, OX11 0QX, UK}
\author{Victor de Haan}\affiliation{BonPhysics Research and Investigations BV, Laan van Heemstede 38, 3297AJ Puttershoek, The Netherlands}

\date{\today}
\hyphenpenalty=800\relax
\exhyphenpenalty=800\relax
\sloppy
\setlength{\parindent}{0pt}
\begin{abstract} 
A coupling between Earth's rotation and orbital angular momentum (OAM), known as the Sagnac effect, is observed in entangled neutrons produced using a spin echo interferometer. After correction for instrument systematics the measured coupling is within 5\% of theory, with an uncertainty of 7.2\%. The OAM in our setup is transverse to the propagation direction and scales linearly with neutron wavelength (4 \AA - 12.75 \AA), so the Sagnac coupling can be varied without mechanically rotating the device, which avoids systematic errors present in previous experiments. The detected transverse OAM of our beam corresponds to $4098\pm295 \ \hbar\mathrm{\AA}^{-1}$, $10^5$ times lower than in previous neutron experiments. This demonstrates the feasibility of using the Sagnac effect to definitively measure neutron OAM and paves the way towards a future observation of the quantum Sagnac effect.
\end{abstract}

\maketitle
\section{Introduction}
In non-inertial frames of reference the laws of nature may appear to work differently, due to additional quasi-forces that arise as a result of the non-uniform motion of the reference frame. George Sagnac's 1913 experiment is a famous example of such an effect. In an attempt to prove the existence of an ether, Sagnac observed a phase shift in his interferometer which depended on the direction in which the interferometer was rotated \cite{Sagnac1913}. The Sagnac effect manifests as an apparent coupling between the rotation frequency of the observer and the angular momentum of the test particle \cite{Post1967}. The famous 1925 experiment by Michelson, Gale and Pearson, was able to observe this coupling in a very large optical interferometer due to the Earth's rotation~\cite{Michelson1925}. Since Sagnac and Michelson made their observations with light, the experimental question remained whether or not matter waves also experience a pseudo-potential in a rotating frame. \myred{In 1965 the effect was first observed in superconducting electrons \cite{Zimmerman1965}.} \myred{Then in 1979, Werner, Staudenmann, and Colella managed to demonstrate that free matter waves are also subject to the Sagnac effect. Specifically neutrons traversing a rotating intereferometer also experience a phase shift proportional to the inner product between the rotation frequency, $\Omega$ and the Orbital Angular Momentum (OAM), $L$ spanned by the neutron's motion in the interferometer ($\Omega\cdot L$) \cite{Werner1979}}. The rotation of the interferometer was realized by the natural rotation of the earth, as was the case in the 1925 Michelson/Gale/Pearson experiment. In addition to neutrons, the Sagnac effect has also been observed in other \myred{free} matter waves, such as electrons \cite{Hasselbach1993} and atoms \cite{Riehle1991,Gautier2022}. In 1988 Mashhoon demonstrated that \myred{applies to the total angular momentum 
\begin{equation}
    \mathbf{\hat{J}}=\mathbf{\hat{L}}+\mathbf{\hat{S}}
\end{equation}
 therefore extrending to the spin angular momentum $\mathbf{\hat{S}}$ of particles as well as their orbital angular momentum \cite{Mashhoon1988}.} Recently the Mashhoon effect has been observed in neutron polarimetry \cite{Demirel2015} and in neutron interferometry \cite{Danner2020}. In these cases the rotating frame was realized by creating a rotating magnetic field in the laboratory frame of the interferometer which, for the projective measurement of the neutron spin employed in the apparatus, can be shown to be equivalent to observing the neutron spin in a rotating frame of reference. 
\par
In parallel with this work on the Mashhoon effect, neutron researchers started to develop methods to produce neutrons with quantized OAM~\cite{Clark2015,Sarenac2019,Geerits2021,Sarenac2022}. \myred{The mathematical form of quantum mechanical OAM is identical to the usual definition of angular momentum in the classical limit $\hat{L}=\hat{r}\times\hat{p}$, with the position and momentum promoted to operators. In addition OAM distinguishes itself from its classical analog by existing only as discrete integer multiples of $\hbar$ upon measurement. Quantum OAM has various expressions in nature, ranging simply from frame dependent "extrinsic" OAM, which is OAM that arises from the choice of reference frame, to frame independent "intrinsic" OAM, such as orbitals of bound particles or the comparatively newer helical/twisted waves. OAM can be considered to be \lq\lq intrinsic \rq\rq if the expectation value of $\hat{L}$ is spatially translation invariant \cite{Berry1998,Neil2002,Geerits2023}. Later we will demonstrate that both extrinsic and intrinsic OAM can be seen as a type of wave structure and that extrinsic states can be expressed as superpositions of intrinsic states. We will also show that extrinsic is not synoymous with not "quantum". Finally we note that intrinsic and extrinsic OAM are parts of the same operator, $\hat{L}$, hence a potential which depends on $\hat{L}$, does not discriminate between intrinsic and extrinsic OAM.} \par
Detection of neutron OAM remains difficult and is the subject of debate and discussion in the scientific literature~\cite{Cappelletti2018,Treimer2023}. Current proposed methods are based on neutron scattering and absorption~\cite{Afanasev2019,Afanasev2021,Jach2022}. The simplicity of the effective potential created by the Sagnac effect, which contains only the OAM operator and an external quantity, could prove useful for the definitive detection of neutron OAM. The neutrons used in the 1975 experiment possessed OAM with respect to the center of the interferometer on the order of $10^9\hbar$. This OAM was detected by the rotation of the earth. \par
\myred{In this paper we present an experiment testing the feasibility of measuring OAM using the Sagnac effect. To this end a neutron interferometer with a precisely calibrated path separation was used. Since, this spatial separation and the neutron momentum are precisely known, it follows that the extrinsic transverse OAM carried by the neutrons is also precisely known. By measuring the resulting Sagnac phase we are able to determine the sensitivity of the setup to any transverse OAM, whether extrinsic or intrinsic. As noted previously potentials depending on $\hat{L}$, do not discriminate between intrinsic and extrinsic OAM, hence if our method successfully detects extrinsic OAM it will also detect intrinsic OAM and vice-versa.}
Compared to the previous neutron Sagnac effect measurement, We improve the angular momentum sensitivity by 5 orders of magnitude. This improvement in sensitivity marks an important step towards observation of the quantized Sagnac effect. In optics this quantized Sagnac effect has been observed using spinning Dove prisms \cite{Courtial1998}. The observation of a quantized energy shift from the Sagnac effect is an attractive method to resolve quantized OAM states in neutrons. We report on an experimental observation of the Sagnac effect in a neutron interferometer which uses microscopic path separation on the order of the transverse coherence of the neutron ($0.001 \mu m - 100 \mu m$) \cite{Rauch1996,Wagh2011,McKay2024}. In addition we demonstrate that the Sagnac effect provides a good method for a basis to definitively detect the OAM of a particle. \myred{In the theory section we show that our setup not only produces the extrinsic transverse OAM, which we measure via the Sagnac effect, but also a much smaller intrinsic longitudinal OAM, that may become accessible if sensitivity is improved. In the discussion we propose to use rotating dove prisms to improve sensitivity such that this longitudinal OAM can be observed.} Our experiment was carried out on the Larmor instrument at the ISIS pulsed neutron source. Larmor is a neutron spin echo type interferometer \cite{Gahler1996}, which employs shaped RF spin flippers to induce horizontal spin and energy dependent path separation \cite{Bouwman2011,Geerits2019}. As a result the spin, energy and path degrees of freedom of the neutron are entangled \cite{Shen2020,Kuhn2021}. \myred{The latter had previously only been possible in perfect crystal neutron interferometry \cite{Hasegawa2010}.}
\section{Theory}
\myred{In this section we distinguish between intrinsic and extrinsic OAM and attempt to express the latter in terms of the former. Next we derive a formula for converting Cartesian position states into cylindrical (extrinsic) OAM states. Finally we will explore the Sagnac effect in neutron spin echo interferometry, SESANS. However first we will start with some general defintions and properties of the OAM operator 
\begin{equation}\label{OAM_Operator}
    \hat{\mathbf{L}}=\hat{\mathbf{r}}\times\hat{\mathbf{p}}
\end{equation}
as shown in \cite{Bliokh2015} in cylindrical coordinates we can focus on the z-component since this is the only component that produces a non-zero expectation value
\begin{equation}\label{OAM_Operator_z}
    \hat{{L}}_z=-\mathrm{i}\hbar(x\frac{\partial}{\partial y}-y\frac{\partial}{\partial x})=-\mathrm{i}\hbar\frac{\partial}{\partial \phi}
\end{equation}
This operator has the Eigenfunctions $e^{\mathrm{i}\ell\phi}$
\begin{equation}\label{Eigenfunctions}
    \hat{L}_z e^{\mathrm{i}\ell\phi}=\ell e^{\mathrm{i}\ell\phi}
\end{equation}
with $\ell$, the OAM mode number, being integer. Since these vortex functions are also Eigenfunctions of the free space Schroedinger equation in cylindrical coordiantes, it follows that any free space wavefunction can be written through superpositions of vortex modes:
\begin{equation}\label{Superposition}
    \psi=\sum_\ell f^\ell(r,z)e^{\mathrm{i}\ell\phi}
\end{equation}
Using this knowledge we can define an OAM distribution function, a probability density function, telling us the chance of finding the wavefunction in the $\ell\mathrm{th}$ mode upon measurement.
\begin{equation}\label{Distribution}
    p[\ell]=\frac{\int \mathrm{d}r\mathrm{d}z r |f^\ell(r,z)|^2}{\sum_\ell \int \mathrm{d}r\mathrm{d}z r |f^\ell(r,z)|^2}
\end{equation}
The $n\mathrm{th}$ moment of the OAM operator can be extracted simply from the OAM distribution function using the relation
\begin{equation}
    <\hat{L}_z^n>=\sum \ell^n p[\ell]   
\end{equation}
For a variety of wavefunctions the first moment $n=1$ (i.e. the expectation value) is spatially translation invariant. In this case one speaks of intrinsic OAM, which is further discussed in the next section.}
\subsection{Intrinsic and Extrinsic OAM}
\myred{Intrinsic OAM distinguishes itself from extrinsic OAM in that the former is invariant under frame translations. It has been shown in multiple publications \cite{Berry1998,Neil2002,Geerits2023}, that this is the case if there is no propagation in the plane in which the OAM is defined. This plane is henceforth referred to as the $x$-$y$ or $r$-$\phi$ plane (i.e. for OAM to be fully intrinsic we require $<k_x>=<k_y>=0$). It can easily be shown that the Eigenstates of the OAM operator in cylindrical coordinates satisfy this condition:
\begin{subequations}
	\begin{eqnarray}
		<k_x>\propto\int \mathrm{d}\phi \ e^{-\mathrm{i}\ell\phi}\frac{\partial}{\partial x} e^{\mathrm{i}\ell\phi}\propto\int \mathrm{d}\phi \ \ell \sin(\phi)=0 \\
		<k_y>\propto\int \mathrm{d}\phi \ e^{-\mathrm{i}\ell\phi} \frac{\partial}{\partial y}e^{\mathrm{i}\ell\phi}\propto\int \mathrm{d}\phi \ \ell \cos(\phi)=0
	\end{eqnarray}
\end{subequations}
noting that $\frac{\partial}{\partial x}=\cos(\phi)\frac{\partial}{\partial r}-\frac{\sin(\phi)}{r}\frac{\partial}{\partial \phi}$ and $\frac{\partial}{\partial y}=\sin(\phi)\frac{\partial}{\partial r}+\frac{\cos(\phi)}{r}\frac{\partial}{\partial \phi}$.
However superpositions of Eigenfunctions do not satisfy this condition:
\begin{subequations}
	\begin{eqnarray}
		<k_x>\propto \int \mathrm{d}\phi \ \sum_{\ell,n} e^{-\mathrm{i}\ell\phi} \frac{\partial}{\partial x} e^{\mathrm{i}n\phi} \\
		<k_y>\propto\int \mathrm{d}\phi \ \sum_{\ell,n} e^{-\mathrm{i}\ell\phi}\frac{\partial}{\partial y}e^{\mathrm{i}n\phi}
	\end{eqnarray}
\end{subequations}
which can be simplified to
\begin{subequations}
	\begin{eqnarray}
		<k_x>\propto \int \mathrm{d}\phi \ \cos(\phi)\sum_{\ell,n} e^{-\mathrm{i}(\ell-n)\phi} \\
		<k_y>\propto\int \mathrm{d}\phi \ \sin(\phi)\sum_{\ell,n} e^{-\mathrm{i}(\ell-n)\phi}
	\end{eqnarray}
\end{subequations}
which are not zero if $\ell-n=\pm 1$. This means that the OAM carried by a wavefunction that consists of a superposition of two or more vortex states with neighboring mode numbers is not entirely intrinsic. We argue that if a state carries extrinsic OAM, this does not necesarrily imply that the state or its OAM is not "quantum". Take the states $\ell$, $\ell+1$ and $\ell + 2$, each of which individually is regarded as a quantum state, the superposition of $\ell$ and $\ell + 2$, carries only intrinsic OAM and is also considered a standard quantum mechanical state. It follows that superpositions that include $\ell + 1$ are also "quantum" and carry extrinsic OAM. 
\begin{figure*}
\includegraphics[width=1\textwidth]{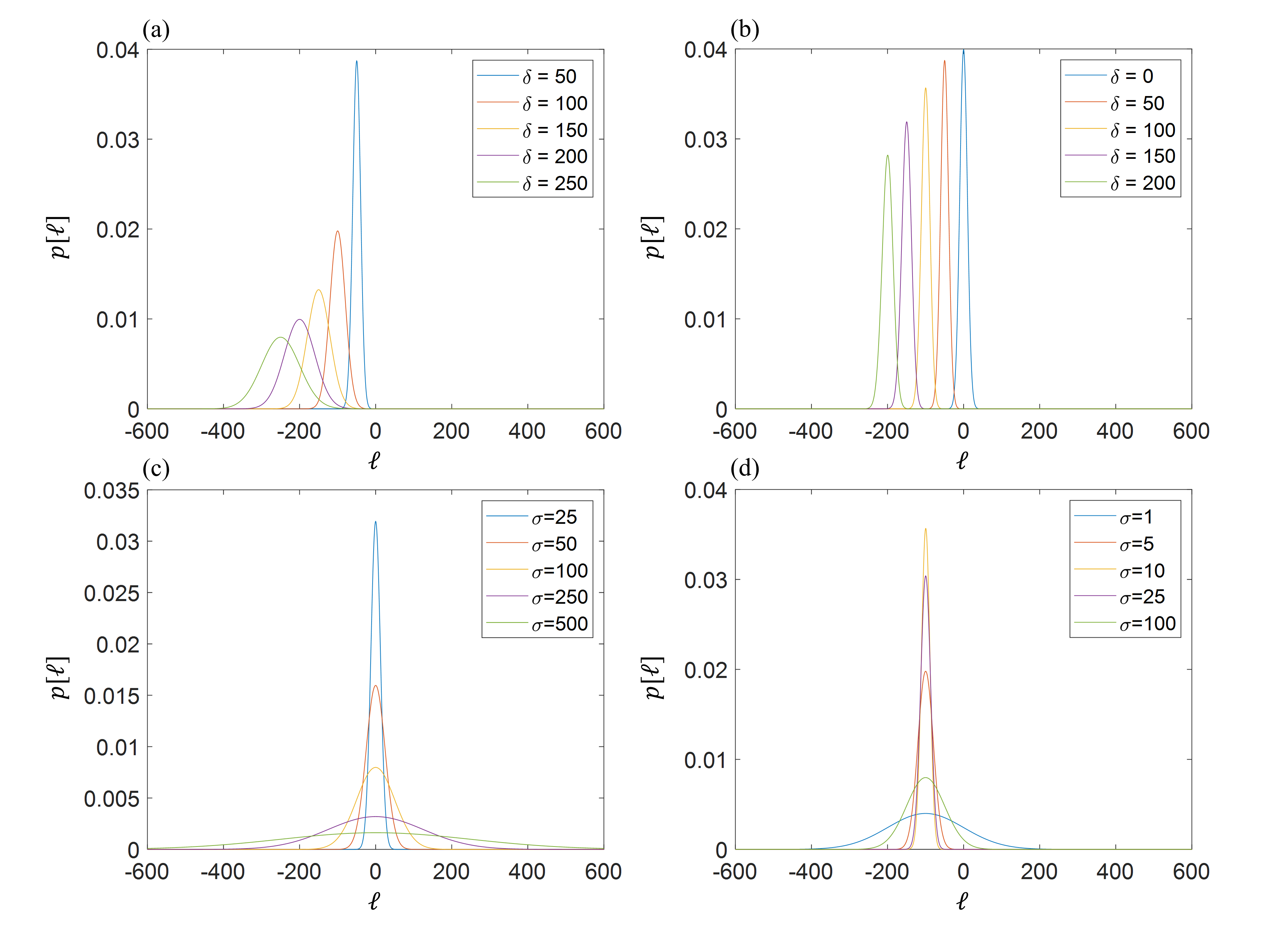}\caption{OAM distribution function  of a planewave with a Gaussian envelope according to equation \ref{PWTOAM} for various different coherence lengths, $\sigma$ and offsets from the cylinder axis $\delta$. In insets (a) and (b) the coherence length is fixed to $5$ and $20$ respectively, while in (c) and (d) $\delta$ has been fixed to $0$ and $100$ respectively. The momentum $k_y$ is $-1$ in all cases.}\label{TWOAM}
\end{figure*}
Extrinsic OAM can emerge from superpositions of intrinsic states and vice versa as shown later. Extrinsic OAM mimics classical AM in that it is coordinate dependent and can be calculated as \cite{Bliokh2015} 
\begin{equation}
    <\hat{\mathbf{L}}_{ext}>=<\hat{\mathbf{r}}>\times<\hat{\mathbf{p}}>=<\hat{\mathbf{L}}>-<\hat{\mathbf{L}}_{int}>
\end{equation}
as one would calculate classical AM. This sometimes leads to the erroneous claim that extrinsic OAM is not related to wave structure. This disregards the fact that both the expectation values $<\hat{\mathbf{r}}>$ and $<\hat{\mathbf{p}}>$ are related to the wave structure. \par
We note that the second moment of the OAM operator 
\begin{equation}\label{OAMSecondMoment}
	\hat{L}_z^2=-\hbar^2 (x^2\frac{\partial^2}{\partial y^2}+y^2\frac{\partial^2}{\partial x^2}-2xy\frac{\partial}{\partial x}\frac{\partial}{\partial y}-y\frac{\partial}{\partial y}-x\frac{\partial}{\partial x})
\end{equation}
is not translation invariant. This can be demonstrated by taking the difference between $\hat{L}_z^2$ and a translated version $\hat{L'}_z^2$ (i.e. $x'\rightarrow x+x_0$ and $y'\rightarrow y+y_0$), which is given by
\begin{equation}\label{DifferenceSecond}
\begin{aligned}
    &\Delta	\hat{L}_z^2=-\hbar^2 ((x_0^2-2xx_0)\frac{\partial^2}{\partial y^2}+(y_0^2-2yy_0)\frac{\partial^2}{\partial x^2}+ \\&2(xy_0+yx_0)\frac{\partial}{\partial x}\frac{\partial}{\partial y}+y_0\frac{\partial}{\partial y}+x_0\frac{\partial}{\partial x})
\end{aligned}
\end{equation}
Each individual term needs to produce a zero expectation value if the second moment of $\hat{L}_z$ is to be translation invariant. For the Eigenfunctions of $\hat{L}_z$ it can be easily shown that the momentum spread operators $\propto\frac{\partial^2}{\partial x^2}$ and $\propto \frac{\partial^2}{\partial y^2}$ produce a non-zero expectation value. For example
\begin{equation}
	<\hat{p}_x^2>=-\int \mathrm{d}\phi e^{-\mathrm{i}\ell\phi}\frac{\partial^2}{\partial x^2}e^{\mathrm{i}\ell\phi}=\pi \ell^2
\end{equation}
It follows that the second moment of the OAM operator is not translation invariant, meaning that the OAM distribution function (eq. \ref{Distribution}) changes when viewing the wavefunction in another frame of reference. We underline the importance of this fact, since many OAM dependent interactions, such as scattering and absorption of twisted neutrons \cite{Afanasev2019,Afanasev2021,Jach2022} do not depend on the expectation value $<\hat{L}_z>$ but rather on the amplitude of the respective mode number. We contrast this with the method described in this paper, which measures the difference between the OAM expectation values of two states. \par
We now examine wavefunctions with extrinsic OAM, and attempt to demonstrate that these can be seen as "partial" vortex states. We start from a wavefunction with purely extrinsic OAM: a planewave with Gaussian envelope propagating offset from the cylinder axis, by $\delta$ ($z$-axis) in the $x$-$y$ plane
\begin{equation}\label{TestWavefunction}
    \psi_t=A e^{-\frac{(x-\delta)^2}{\sigma^2}-\frac{y^2}{\sigma^2}}e^{ik_y y}
\end{equation}
$\sigma$ denotes the coherence length \cite{Rauch1996,McKay2024} of the wavepacket.
It is clear that the total and extrinsic OAM are equal and transverse to the propagation direction
\begin{equation}
	<\psi_t|\hat{L}_{z}|\psi_t>=<\psi_t|\hat{L}_{z,ext}|\psi_t>=\hbar \delta k_y \hat{z}
\end{equation}
it follows that the intrinsic OAM component is zero. As stated earlier vortex modes form a complete basis, hence we can expand \ref{TestWavefunction} in terms of vortex states. This can be done using the Jacobi-Anger expansion \cite{Abramowitz}
\begin{equation}
	\psi_t=Ae^{-\frac{r^2+\delta^2}{\sigma}}\sum_{m,n} \mathrm{i}^{m} J_m(-2\mathrm{i}\frac{\delta r}{\sigma^2})J_n(k_y r)e^{\mathrm{i}(m+n)\phi}
\end{equation}
The $\ell\mathrm{th}$ mode of this wavefunction is therefore given by
\begin{equation}
	\psi_t^\ell(r)=Ae^{-\frac{r^2+\delta^2}{\sigma^2}}\sum_{m} \mathrm{i}^{m} J_{m}(-2\mathrm{i}\frac{\delta r}{\sigma^2})J_{\ell-m}(k_y r)
\end{equation}
which can be simplified using Grafs addition theorem for Bessel functions \cite{Watson1944}
\begin{equation}\label{Vortex_PW}
	\psi_t^\ell(r)=Ae^{-\frac{r^2+\delta^2}{\sigma^2}} e^{\mathrm{i} \ell \alpha} J_\ell (k'r)
\end{equation}
with $k'=\sqrt{k_y^2 - 4\frac{\delta^2}{\sigma^4}}$ and $\alpha=\sin^{-1}(-2\mathrm{i}\delta/\sigma^2k')=\cos^{-1}(k_y/k')$. Equation \ref{Vortex_PW}, tells us which vortex states $\psi_t^\ell$, each individually carrying intrinsic OAM, make up the Gaussian planewave carrying exclusively extrinsic OAM. Finally the OAM distribution function can be determined according to equation \ref{Distribution}. Which becomes a standard Hankel transform given in \cite{Bateman1954} with the result
\begin{equation}\label{PWTOAM}
	p[\ell]= \frac{A^2\sigma^2 }{4}e^{\mathrm{i}\ell(\alpha-\alpha^*)}e^{-\frac{\sigma^2 k'^2}{4}-\frac{2\delta^2}{\sigma^2}} I_\ell(\frac{\sigma^2|k'|^2}{4})
\end{equation}
For the sake of illustration $p[\ell]$ is shown for a few combinations of $\sigma$ and $\delta$ in figure \ref{TWOAM}
One can see that the central mode is always given by $\ell \approx k_y\delta$ as expected and that point like particles far away from the cylinder axis (i.e. particles with short coherence lengths and $\delta>\sigma$) have large OAM spreads. For illustration in figure \ref{Comparison} we compare the main cylinder mode of a Gaussian planewave given in equation \ref{Vortex_PW} with the Gaussian planewave.
\begin{figure*}
\includegraphics[width=1\textwidth]{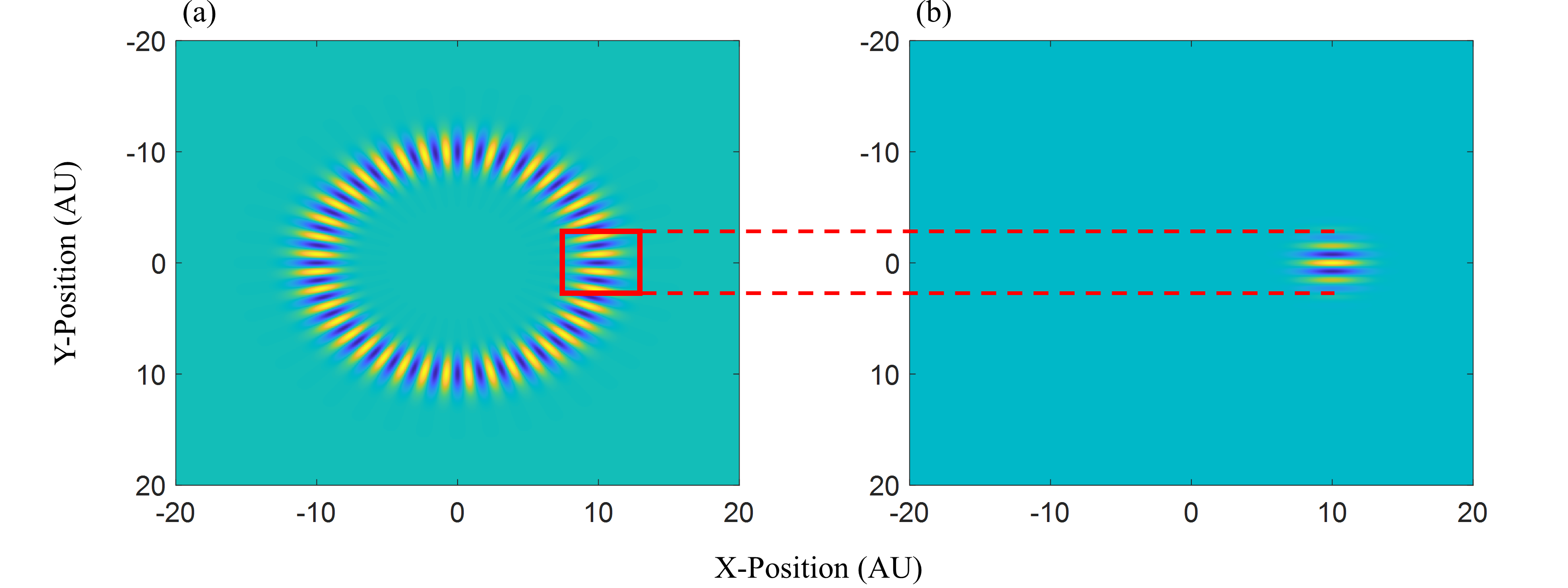}\caption{Comparison of the real part of a planewave with a Gaussian envelope as described in equation \ref{TestWavefunction}, with $k_y=4$, $\delta=10$ and $\sigma=2$ (b) and the main vortex mode ($\ell=40$) contributing to the Gaussian planewave according to \ref{Vortex_PW} (a). A red rectangle indicates the similarity between the vortex state and the offset Gaussian planewave.}\label{Comparison}
\end{figure*}
While figure \ref{TWOAM}, shows clearly that many vortex modes can be used to contruct a wavefunction with extrinsic OAM, figure \ref{Comparison}, demonstrates that the inverse is also possible. Using a superposition of Gaussian planewaves carrying extrinsic OAM one could approximate the vortex state shown in figure \ref{Comparison}. Most importantly both figures demonstrate qualitatively what we have previously quantitatively derived, extrinsic OAM can be seen as emerging from the wave structure of many intrinsic vortex states and vice-versa. Finally we wish to underline the key points to take away from this section (1) intrinsic and extrinsic OAM are intricately related, such that they can be converted into each other, (2) both extrinsic and intrinsic OAM can be understood as quantum phenomena and (3) if the reference frame is fixed, both extrinsic and intrinsic OAM will produce the same observable phenomena.}
\begin{figure*}
	\includegraphics[width=18cm]{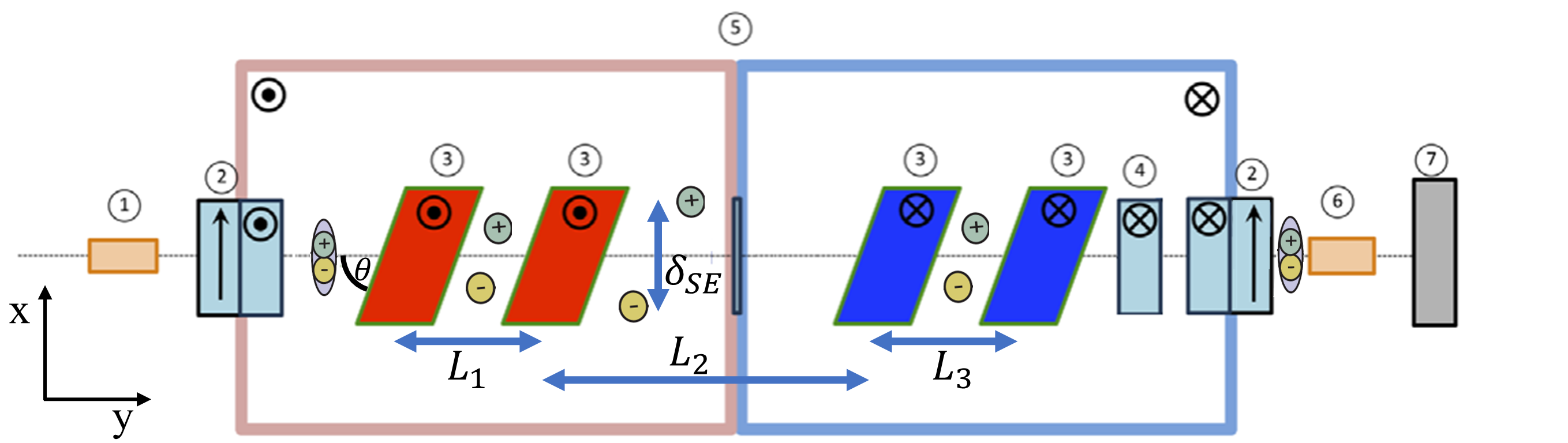}
	\caption{Schematic of the Larmor neutron spin echo interferometer, the propagation of the + and - spin through the device and the neutron optical components: (1) polarizer (2) adiabatic $\pi/2$ rotator (v-coil), (3) RF spin flipper with tilted field region, (4) ramped $\pi/2$ rotator (5) guide field, (6) spin analyzer and (7) detector.}\label{interferometer}
	\label{Setup}
\end{figure*}
\subsection{The Sagnac Effect in SESANS}
\myred{In this section, we derive the wavefunction produced by a SESANS type interferometer, determine its OAM and explore how the Sagnac effect couples to this OAM.}
In a rotating frame of reference, particles experience a pseudo potential proportional to the rotation rate of the frame and the OAM possessed by the particle around the axis of rotation
\begin{equation}\label{SagnacPotential}
    \hat{V} \propto \Omega \cdot \hat{L}
\end{equation}
\myred{Intuitively, such a potential allows us to measure the OAM component of a wavefunction parallel to the axis of rotation. \par}
\myred{Our setup is shown in figure \ref{Setup}. It is clear that the wavefunction of each spin state can be approximated by the Gaussian planewave we used previously to explore extrinsic OAM (eq. \ref{TestWavefunction})
\begin{equation}\label{Wavefunction}
    \psi_\pm=\frac{A}{\sqrt{2}} e^{\pm\mathrm{i}\beta/2} e^{-\frac{(x\pm\delta_{SE}/2)^2-y^2-z^2}{\sigma^2}}e^{ik_y y}
\end{equation}
the $z$-component has been added to the expression so that we can later also analyze the longitudinal OAM of this wavefunction. The $\pm$ refers to the spin state either being aligned with the $z$-axis $+$ or anti-aligned $-$. $\beta$ is an arbitrary phase between the up and down spin state. As shown previously the average OAM mode transverse to the propagation direction is
\begin{equation}\label{Eigenvalue}
    \ell_\pm=\pm \frac{\delta_{SE}|k|}{2}
\end{equation}
Usually the state in the interferometer is expressed as a spin-path entangled state. The path states are often characterized by the spin echo length $\delta_{SE}$. As we have seen this degree of freedom also characterizes the transverse OAM of the wavefunction. It follows that the two path states, defined by the spin echo length, can also be described by two OAM states. Thus, for the purposes of this experiment, we describe our state not as spin-path entangled, but as spin-orbit entangled, such that the state in the interferometer can be described as
\begin{equation}\label{Entangled}        \ket{\psi}=^{i\beta/2}\ket{\ell_+}\ket{+}+e^{-i\beta/2}\ket{\ell_-}\ket{-}
\end{equation}
We note that state preparation and measurement are not instantaneous. Between the first and second RF flipper spin echo length and therefore transverse OAM is linearly increased, while between the third and fourth RF flipper, the spin echo length and transverse OAM are reduced back to zero. The Sagnac effect will obviously also act on these intermediate states. This is taken into account by allowing for an OAM which depends on the $y$-coordinate.} The precession frequency between the two states due to the Sagnac effect follows from eq. \ref{SagnacPotential} and \ref{Eigenvalue} and is given by
\begin{equation}
    \delta\omega_s=[\ell_+-\ell_-]\Omega \sin(\Lambda)
\end{equation}
with $\Lambda$ the latitude of the interferometer. \myred{This expression demonstrates one of the advantages of using the Sagnac effect to measure OAM: the precession frequency depends only on the OAM difference between the two states. This difference is invariant under spatial translations of the frame of reference, hence intrisicality/extrinsicality becomes an irrelevant detail.}
Integrating this precession frequency over the length of the instrument leads to the Sagnac phase shift
\begin{equation}\label{OAMPhase}
    \delta\phi_s = \int \ dt \ \delta\omega_s = \frac{m\Omega\Delta\ell}{\hbar|k|}\sin(\Lambda)[L_1+L_3+2L_2],
\end{equation}
which can be reduced to the result shown in \cite{Werner1979,Haan2014}, for a horizontal interferometer
\begin{equation}\label{AreaPhase}
    \delta\phi_s=\frac{2mA\Omega}{\hbar}\sin(\Lambda)
\end{equation}
with A the area spanned by the two paths of the interferometer. The area, illustrated in figure \ref{interferometer}, is given by
\begin{equation}\label{Area}
    A=\delta_{SE}[\frac{L_1+L_3}{2}+L_2].
\end{equation}
As will be shown in the next section the spin echo length, $\delta_{SE}$ is proportional to wavelength squared in spin echo interferometers.
Hence to vary the area of the interferometer and therefore strength of the Sagnac effect, we may simply vary the wavelength of incident neutrons, whereas the 1979 perfect crystal experiment required physical rotation of the interferometer, which may induce systematic errors. 
\myred{We now characterize the longitudinal and intrinsic OAM of the wavefunction produced in a spin echo interferometer. The most interesting case to look at is when the wavefunction (eq. \ref{Wavefunction}) is post-selected such that spin information is erased. We will look only at the transverse part of the wavefunction since only this is relevant to the longitudinal OAM
\begin{equation}\label{WavefunctionLongitudinal}
    \psi_T=\frac{A}{\sqrt{2}} [e^{\mathrm{i}\beta/2} e^{-\frac{(x+\delta_{SE}/2)^2-z^2}{\sigma^2}}+e^{-\mathrm{i}\beta/2} e^{-\frac{(x-\delta_{SE}/2)^2-z^2}{\sigma^2}}]
\end{equation}
We wish to derive the individual vortex modes, equivalent to eq. \ref{Vortex_PW}, in the longitudinal direction and the respective OAM distribution function. This is relatively simple if we realize that both terms in eq. \ref{WavefunctionLongitudinal} are translated versions of the wavefunction \ref{TestWavefunction} in the special case where $k_y=0$. Hence we can reuse our results eq. \ref{Vortex_PW} and eq. \ref{PWTOAM} to characterize longitudinal OAM of the wavefunction eq. \ref{WavefunctionLongitudinal}. The vortex modes are given by
\begin{equation}\label{Vortex_PW2}
	\psi_{t,2}^\ell(r)=\sqrt{2}Ae^{-\frac{r^2+\delta^2}{\sigma^2}} J_\ell (k'r) \cos(\ell\alpha + \beta/2)
\end{equation}
and the OAM distribution function is given by
\begin{equation}\label{PWTOAM2}
	p[\ell]= \frac{A^2\sigma^2 }{2}e^{-\frac{\sigma^2 k'^2}{4}-\frac{2\delta^2}{\sigma^2}} I_\ell(\frac{\sigma^2|k'|^2}{4})|\cos|^2(\ell\alpha+\beta/2)
\end{equation}
with $\delta=\delta_{SE}/2$, $k'=2\mathrm{i}\frac{\delta}{\sigma^2}$ and $\alpha=\pi/2$. The most interesting cases occur when $\beta=0$, in which case all odd vortex modes are eliminated and $\beta=\pi$, where all even modes are eliminated. A few plots of eq. \ref{PWTOAM2} are shown in figure \ref{EvenOdd}, for $\beta=0$ and $\beta=\pi$ and various coherence length $\sigma$, the splitting parameter $\delta$ is left constant. One can see that once $\sigma$ exceeds $\delta$, that is to say if the splitting parameter is much smaller than the coherence length, only the lowest order vortex modes contribute to the overall wavefunction.
\begin{figure*}
	\includegraphics[width=18cm]{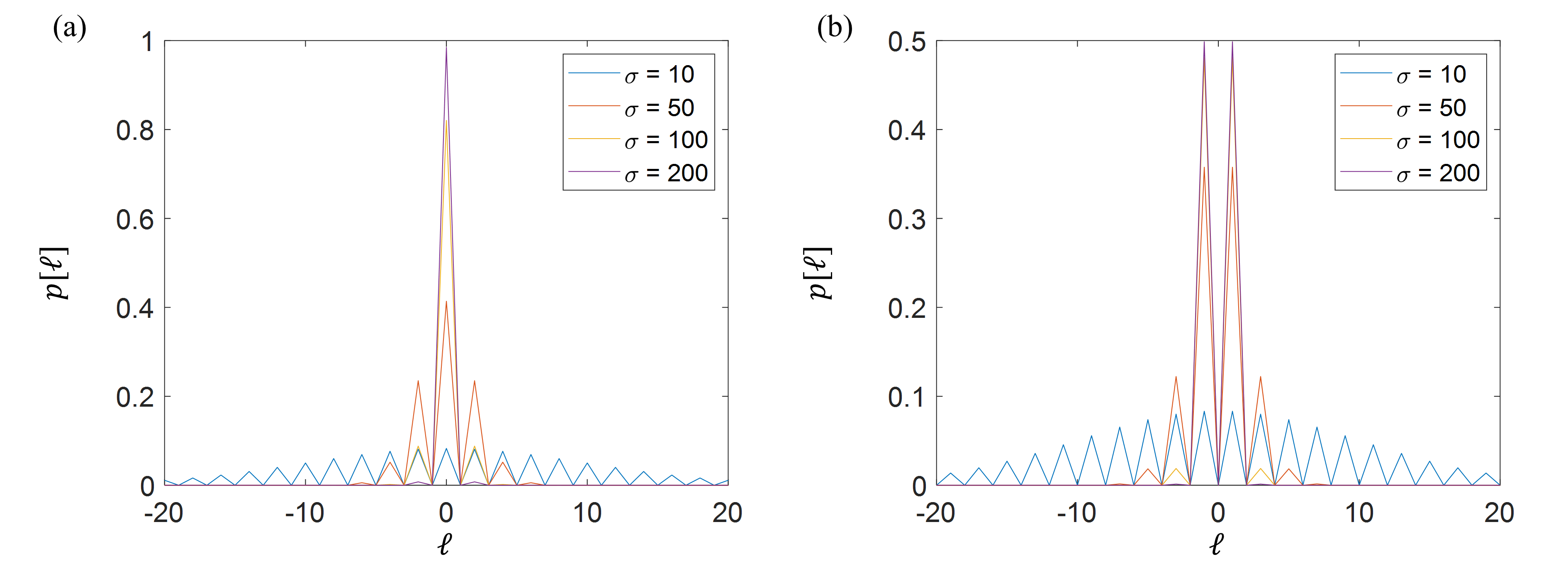}
	\caption{OAM distribution functions describing the longitudinal vortex modes carried by the wavefunction eq. \ref{WavefunctionLongitudinal} in the cases (a) $\beta=0$, where only even vortex modes contribute, and (b) $\beta=\pi$, where only odd vortex modes contribute, for various different coherence lengths $\sigma$. The separation parameter is left constant with $\delta=100$. As the coherence length increases relative to the separation constant the number of contributing higher order modes decreases, until the limiting case where only the lowest order modes play a role (i.e. for $\beta=0$ $\ell=0$ and for $\beta=\pi$ $\ell=\pm 1$}\label{EvenOdd}
	\label{EvenOdd}
\end{figure*}
This is particularly interesting in the $\beta=\pi$ case, since for $\delta<\sigma$, the state approximates a superposition of $\ell=1$ and $\ell=-1$. In analogy to spin optics we refer to such a state as a linearly polarized OAM state, since it consists of an equal superposition of the clockwise and counter-clockwise rotating states. As shown before such a state carries intrinsic OAM, as the mode numbers are not neighbors. Though the OAM expectation value of such a state is zero, it distinguishes itself from the $\ell=0$ state, as in theory it produces a non-zero Sagnac phase, given sufficient measurement resolution and is expected to interact differently with matter \cite{Afanasev2019,Afanasev2021,Jach2022}. Since current measurement resolution is insufficient, this paper is only concerned with observing the Sagnac effect due to the transverse extrinsic OAM of the wavefunction.} \par
In equation \ref{OAMPhase} we put the acquired Sagnac phase in a form that shows that one can extract the difference between the quantum numbers for two OAM states, assuming instrument parameters and the rotation rate are well known. Hence, using the Sagnac effect can be a good relative measure to determine the OAM of a beam.
\begin{figure*}
	\includegraphics[width=18cm]{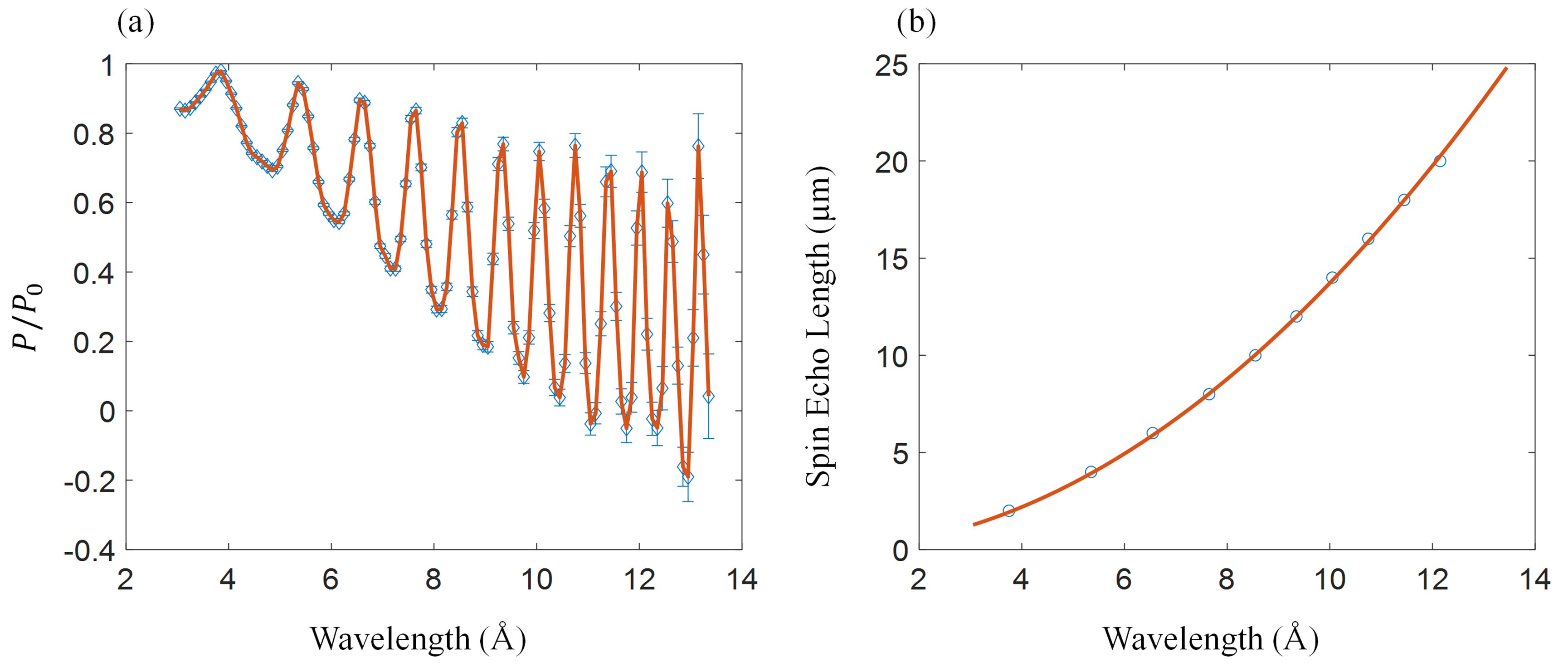}
	\caption{Calibration curves of the instrument produced using a 2 µm period silicon grating. (a) Plot of the normalized spin-echo polarization against wavelength. The n$^{th}$ peak corresponds to a spin echo length equal to n times the grating period. The wavelength and corresponding spin echo length is extracted and plotted in the next panel (b). The quadratic fit is drawn in red.}\label{calibration}
\end{figure*}
\section{Methods}
Our measurements were carried out on the Larmor instrument at the ISIS pulsed neutron source \cite{LarmorData}. Larmor is a SANS instrument with a versatile neutron resonant spin echo toolbox, based on four gradient radio frequency spin flippers with shaped poleshoes, capable of performing inelastic techniques such as Modulated Instensity Emerging from Zero Effort (MIEZE) \cite{Geerits2019} and Spin Echo (Modulated) Small Angle Neutron Scattering (SE(M)SANS) \cite{Rekveldt1996,Li2016,Shen2020,Kuhn2021}. Our experiment makes use of the SESANS mode of the instrument, which employs magnetic refraction from a tilted field region to induce spin dependent spatial splitting and a second oppositely polarized field region to recombine the two split states (see figure \ref{interferometer}). The path separation, also known as the spin echo length is proportional to the wavelength squared
\begin{equation}\label{dse}
    \delta_{SE}=\frac{m\gamma B_0 L_1 \cot(\theta)}{2\pi^2 \hbar } \lambda^2=c_{SE}\lambda^2
\end{equation}
with $\gamma$ the gyromagnetic ratio of the neutron, $\lambda$ the neutron wavelength, $m$ the neutron mass, $B_0$ the magnitude of the static magnetic field inside the RF flippers and $\theta$, the angle between the incident beam and the magnetic field region. The constant $c_{SE}$ which summarizes all instrument parameters and constants, is called the spin echo constant.
This interferometric technique has previously been used to revisit the Colella-Overhauser-Werner experiment \cite{Haan2014}, put limits on exotic spin-gravity couplings \cite{Parnell2020} and measure Bell inequalities with neutrons \cite{Shen2020,Kuhn2021}. \\
Since Larmor uses spin dependent refraction to realise the interferometer, the path and spin states of the neutron are coupled (i.e. mode entangled). Hence any path dependent phase shift is projected onto the spin and vice-versa. As pointed out previously the path and OAM degree of freedom are related, hence the path state and also path phases, may also be described as orbit states/phases. In SESANS the spin is usually prepared, along the x-axis, orthogonal to the beam propagation and $B_0$ direction. The expectation value of the spin, also called polarisation, is usually also measured along the x-direction, leading to a polarisation of
\begin{equation}
    P_x=P_0 \cos(\Delta\phi(\lambda))
\end{equation}
with $P_0=\sqrt{P_x^2+P_y^2+P_z^2}$ and $\Delta\phi(\lambda)$ a polynomial in $\lambda$
\begin{equation} \Delta\phi(\lambda)=a_0+a_1\lambda+a_2\lambda^2+O(\lambda^3).
\end{equation}
One may independently control the $a_0$ term by means of a ramped precession field, with $B(t)\propto 1/t$, which ensures that the spin of each wavelength on a ToF source is rotated by the same angle \cite{Nilsen2017,Cassella2019}. By setting $a_0$ equal to $\pi/2$ we effectively change the measurement direction to along the y-axis, the propagation direction. Hence the measured polarisation becomes
\begin{equation}
    P_y=P_0 \sin(\Delta\phi'(\lambda)),
\end{equation}
which for small $\Delta\phi'$ may be linearized. \myred{For more details on the pre- and post-selection in SESANS type interferometers we refer to \cite{Lu2020}} To remove the scaling factor $P_0$, we may normalise $P_y$ by $P_x$. This normalized polarization still has a simple and accurate linearization for small $\Delta\phi$
\begin{equation}\label{fitfunction}
    \frac{P_y}{P_x}\approx \epsilon + a_1\lambda + a_2\lambda^2 
\end{equation}
with $\epsilon$ any imprecision in the quality of the $\pi/2$ rotation provided by the ramped precession field.
We can estimate the second order parameter, due to the Sagnac effect using equations \ref{AreaPhase} and \ref{dse}
\begin{equation}\label{estimate}
    a_2=c_{SE} \frac{m \Omega}{\hbar}\sin(\Lambda)[L_1+L_3+2L_2].
\end{equation}
However additional perturbations arising from imperfections in the instrument can occur which can affect the magnitude of the second order term. Most notably a slight change in the precession plane can occur if an imperfect spin optical component introduces an unintended low probability spin flip. Components that may be suspected to introduce such an effect are primarily those which use adiabatic field changes to effect a spin rotation, for example v-coils (item 2 in figure \ref{Setup}) and adiabatic RF flippers (item 3 in figure \ref{Setup}), since adiabatic spin flip probabilities in both of these components can be described as
\begin{equation}\label{AdiabaticWobble}
    \rho \approx A_i\lambda^2\cos^2(k_i\lambda+\alpha)
\end{equation}
assuming only slight imperfection \cite{Grigoriev2001,Kraan2003} (i.e. low spin flip probability). As a result the precession plane will appear to oscillate with an amplitude proportional to $\lambda^2$ and in addition this effect will produce an abberation on $a_2$, since the $\cos^2$ component in equation \ref{AdiabaticWobble} produces a constant offset (\myred{$A\cos^2(x)=A/2(1+\cos(2x))$}). This offset can of course be isolated and subtracted from $a_2$, by measuring the amplitude of the precession plane oscillation. Since this effect is small it has not been relevant to measurements conducted with Larmor before, however our experiment has sufficient sensitivity to uncover this systematic. \par
Measurements of the Sagnac phase were conducted using a poleshoe angle of $\theta=40$ degrees and at an RF frequency of 2 MHz corresponding to a magnetic field strength of 68.6 mT. These consisted of polarization measurements with the ramped $\pi/2$ rotator turned off and on with both polarities (corresponding to a $\pi/2$ or $-\pi/2$ rotation). In addition with the ramped $\pi/2$ rotator turned off a calibration measurement was carried out, in which a 2 micron silicon grating was inserted into the sample position of the instrument. The resulting correlation function, shown in figure \ref{calibration} (a), allows us to experimentally determine the proportionality constant between the spin echo length and the wavelength squared, which is essential to estimate $a_2$. Since the n$^{th}$ peak in figure \ref{calibration} (a) corresponds to a spin echo length of n times the grating period, one can extract the relationship between spin echo length and wavelength, shown in figure \ref{calibration} (b). By applying a quadratic fit one finds the spin echo constant, $c_{SE}$, to be equal to $0.137 \mu m \mathrm{\AA}^{-2}$. Using this and equation \ref{estimate} it follows that for the instrument settings used in this experiment the Sagnac constant $a_2$, should be equal to $-1.15\times 10^{-3} \mathrm{\AA}^{-2}$. Equations \ref{Eigenvalue} or \ref{OAMPhase}, show that the difference between OAM states scales linearly with $\lambda$ (i.e. $\delta\ell=c_{OAM}\lambda$). Using equation \ref{OAMPhase} we can express $c_{OAM}$ in terms of $a_2$ 
\begin{equation}\label{OAMCon}
    c_{OAM}=\frac{2\pi \hbar a_2}{m\Omega\sin(\Lambda)[L_1+L_3+2L_2]}
\end{equation}
which for our estimated value of $a_2$ is equal to $-8.62\times10^3\mathrm{\AA}^{-1}$. Since according to equation \ref{Eigenvalue} $\ell_+$ and $\ell_-$ are equal in magnitude it follows that the OAM of each state scales with $\ell_\pm=\pm \frac{1}{2} c_{OAM}\lambda$.
\section{Results}
\begin{figure}
	\includegraphics[width=9cm]{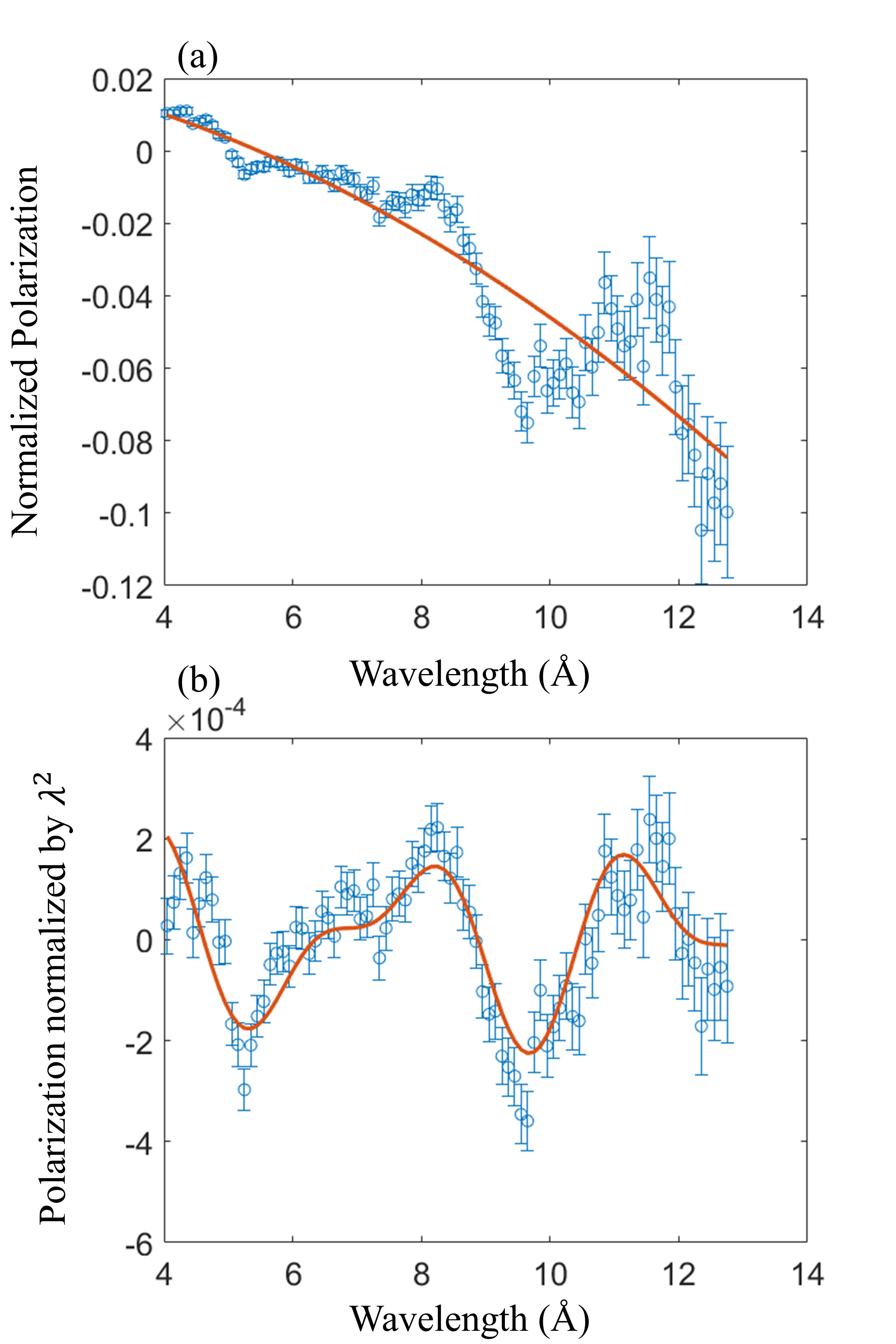}
	\caption{(a) Typical normalized polarization for the 40 degree poleshoe setting, in which the Sagnac phase shift is expected (blue). A quadratic fit is shown in red. The quadratic fit can be subtracted from the data to isolate the oscillations (b) Since the amplitude of the oscillations scale with wavelength squared, we divide these by $\lambda^2$. A fit consisting of two sines is shown in red.}\label{Typical}
\end{figure}
\begin{figure*}
	\includegraphics[width=18cm]{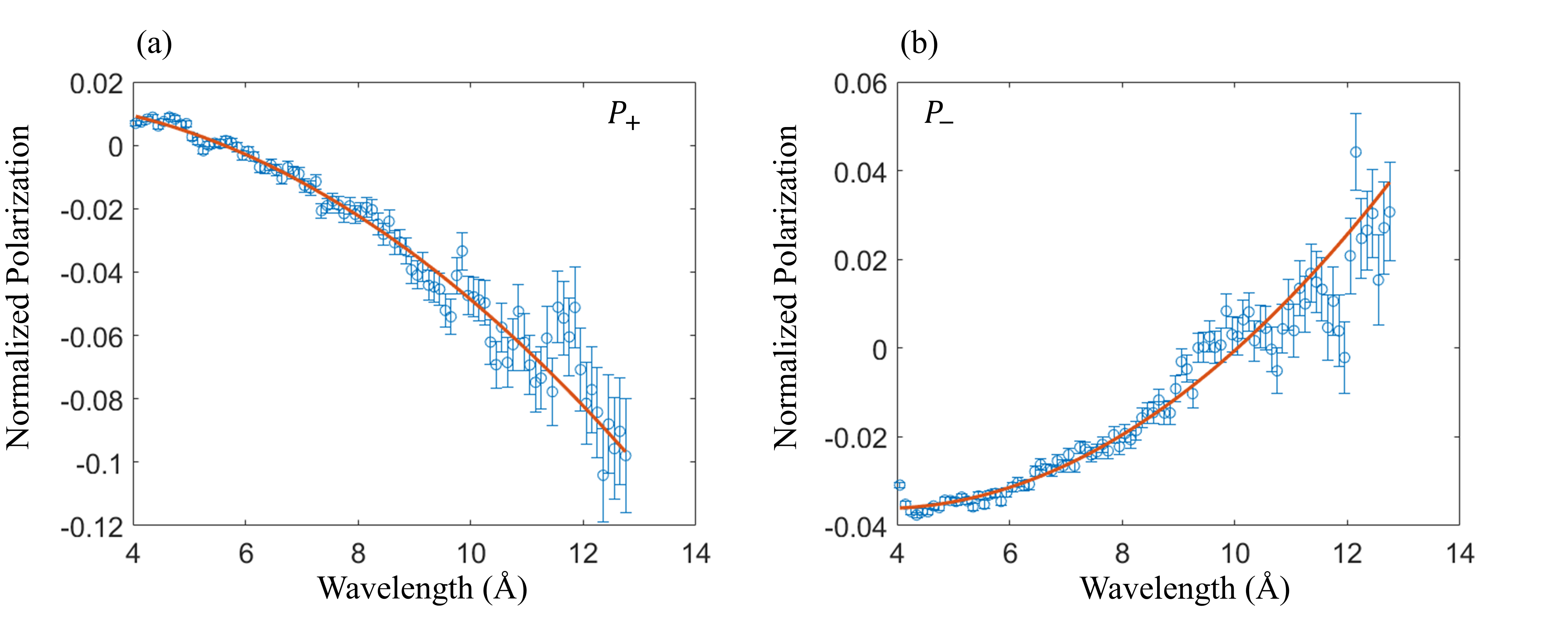}
	\caption{Plots of the normalized and corrected polarization in blue for (a) positive polarity and (b) negative $\pi/2$ rotator polarity. The quadratic fits are plotted in red. The errorbar introduced by subtracting the sinusoidal fit is negligible.}\label{Dewiggled}
\end{figure*}
\begin{figure}
	\includegraphics[width=9cm]{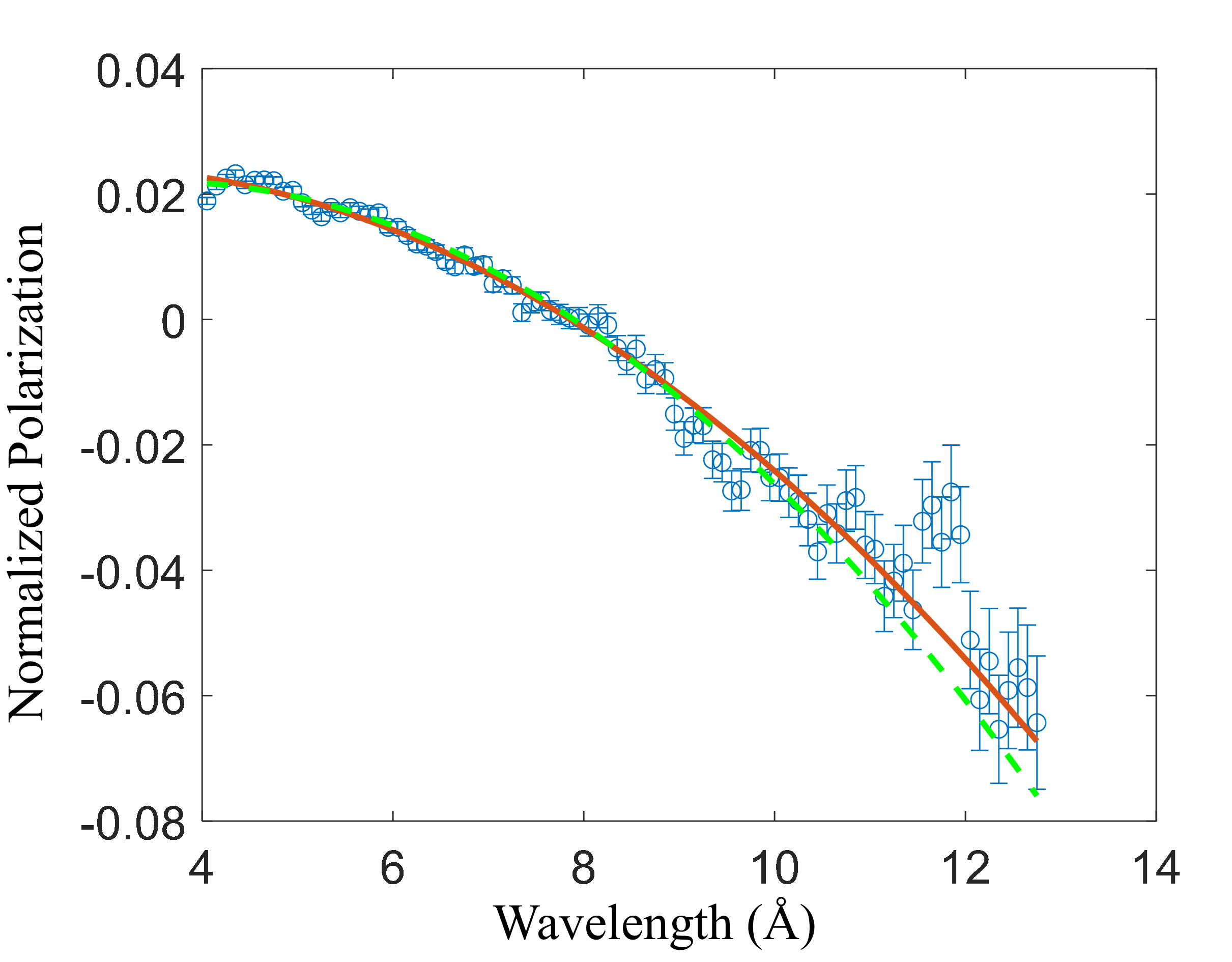}
	\caption{Averaged normalized polarization of all data, according to equation \ref{aggregate}, shown in blue. The quadratic fit is shown as a red solid line. While a fit containing the exact theoretical expectation value for the second order term due to the Sagnac effect is shown as a green dashed line.}\label{Final}
\end{figure}

The normalized polarization, for the positive $\pi/2$ rotator polarity, is shown in figure \ref{Typical} (a). This represents a typical spin-echo curve obtained from this experiment. It is clear that the raw data contains oscillations which increase in amplitude with wavelength squared, analogous to the precession plane oscillation described in the previous section (see eq. \ref{AdiabaticWobble}). By subtracting a quadratic fit from the data and dividing the result by $\lambda^2$, the oscillations can be isolated (see figure \ref{Typical} (b)). 
\begin{table}
    \centering
    \begin{tabular}{|c|c|c|c|} \hline 
         &$a_2 \ (\mathrm{\AA^{-2}})\times 10^3$  &$|A_1| \ (\mathrm{\AA^{-2}})\times 10^5$  &$|A_2| \ (\mathrm{\AA^{-2}})\times 10^5$\\ \hline 
         $P_{+}$ &$-0.891\pm 0.0853 $  &$14.4\pm 2.31$  &$8.22\pm 2.24$ \\ \hline 
         $P_{-}$&$0.898\pm 0.0739$  &$8.88\pm 1.85$  &$6.23\pm 1.83$ \\ \hline
    \end{tabular}
    \begin{tabular}{|c|}\hline
         $a_2^{S}=(-0.894\pm 0.0564)\times 10^{-3}\mathrm{\AA}^{-2}$ $c_{OAM}=-6767\pm 427\mathrm{\AA}^{-1}$ \\ \hline
    \end{tabular}
    \caption{Table containing the second order fit parameters, $a_2$ and their respective standard deviations, for both coil polarities and the amplitudes of the oscillations found in the data with their respective errors. The final estimate for the second order parameter due to the Sagnac effect, $a_2^S$, which is calculated using equation \ref{SagnacParameter}, is shown at the bottom, in addition to the OAM proportionality constant (see equation \ref{OAMCon}).}
    \label{FitValues}
\end{table}
It can be shown that the abberation consists of two oscillations with frequencies $k_2\approx 2k_1$. As pointed out previously it is important to correct for these oscillations, since in addition to improving the overall fit quality, the amplitude information is necessary to correct for a systematic error coming from imperfections of the instrument, hence both amplitudes are listed in table \ref{FitValues}. The data is corrected by fitting two sine waves to the oscillations and subtracting said fit multiplied by $\lambda^2$ from the data. The corrected data for both $\pi/2$ coil settings is shown in figure \ref{Dewiggled}. Quadratic fits using a weighted least squares method are applied to the corrected data. The weights are given by the inverted variance. The second order fit parameters are illustrated in table \ref{FitValues}. 
The first estimate for the second order parameter $a_2^S$, due to the Sagnac effect is obtained using the following
\begin{equation} \label{SagnacParameter}
    a_2^S=\frac{a_2^{+}-a_2^{-}}{2}
\end{equation}
Alternatively the corrected data may be aggregated according to a similar formula
\begin{equation}\label{aggregate}
    P_S=\frac{P_{+}-P_{-}}{2}
\end{equation}
and a weighted least squares quadratic fit is applied to this result. $P_S$ is shown in figure \ref{Final}.
The resulting second order fit parameter is $(-0.899\pm 0.0631)\times10^{-3}$, which is in good agreement with the value $a_2^S$ shown in table \ref{FitValues}. For comparison figure \ref{Final} contains a fit in green which uses the theoretically exact value for the second order parameter, while the zeroth and first order coefficients are determined via least squares regression. 
\section{Discussion}
Figure \ref{Final} indicates a good agreement between our calculated Sagnac parameter of $-1.15\times10^{-3} \mathrm{\AA}^{-2}$ and the measured parameter, however the exact fit parameters shown in table \ref{FitValues}, indicate that for a single $\pi/2$ rotator polarity, the measured value differs from our calculation by roughly $3\sigma$. Furthermore the averaged fit parameter $a_2^S$, differs from theory by $4.3\sigma$, indicating a high likelihood of a systematic perturbation. In this section we correct this perturbation to the second order fit parameter, assuming it is caused by imperfections in the adiabatic spin optical components. Both the oscillations shown in figure \ref{Typical} and the large $\propto 4\sigma$ deviation between data indicate that such imperfections are likely. As indicated earlier at low efficiencies adiabatic spin flip probabilities scale with $\lambda^2$, consistent with our observation. Therefore we postulate that the oscillations shown in figure \ref{Typical} (b) arise due to an oscillation of the precession plane of the form
\begin{equation}\label{Correction}
    P=2\lambda^2[ |A_1|\cos^2(k_1\lambda+\phi_1)+|A_2|\cos^2(k_2 \lambda+\phi_2)]
\end{equation}
similar to equation \ref{AdiabaticWobble}, which introduces a systematic to the second order fit parameter of $\pm[|A_1|+|A_2|]$. Where the sign is determine by the polarity of the $\pi/2$ rotator. The corrected second order parameters, $\Bar{a}_2$, using the amplitudes shown in table \ref{FitValues} are shown in the table \ref{corrected}.
\begin{table}
    \centering
    \begin{tabular}{|c|c|c|c|} \hline 
         &$\Bar{P}_+$  &$\Bar{P}_-$  &$\frac{\Bar{P}_+-\Bar{P}_-}{2}$ \\ \hline 
         $\Bar{a}_2 (\mathrm{\AA}^{-2})\times 10^3$ &$-1.117 \pm 0.121$  &$1.049\pm 0.098$  &$-1.083\pm 0.078$ \\ \hline 
         $\Bar{c}_{OAM} (\mathrm{\AA}^{-1})$ &$-8454 \pm 916$  &$7940\pm 742$  &$-8197\pm 590$ \\ \hline
    \end{tabular}
    \caption{Corrected estimates for the second order parameter due to the Sagnac effect, $\Bar{a}_2$, and their respective errors for both $\pi/2$ coil polarities and their average, in addition to the OAM proportionality constant, $\Bar{c}_{OAM}$ for each setting. For comparison our theoretical estimate for $a_2$ is $-1.15 \times 10^{-3} \mathrm{\AA}^{-2}$ and for $c_{OAM}$ is $-8.62 \times 10^{3} \mathrm{\AA}^{-1}$}
    \label{corrected}
\end{table}
After applying this correction the average estimated second order parameter due to the Sagnac effect is $(-1.083\pm0.078)\times10^{-3}$, which is within $1\sigma$ of the expected theoretical value. This corresponds to an OAM proportionality constant of $-8197\pm 590$ units of $\hbar/\mathrm{\AA}$ according to equation \ref{OAMCon}. From this the \myred{average} OAM Eigenvalues $\ell_\pm$ of the two path states can be extracted: $\ell_\pm=\pm 4098 \pm 295 \hbar/\mathrm{\AA}\cdot\lambda$. This can be compared to the results of our calibration measurement (figure \ref{calibration}), which, based on equation \ref{Eigenvalue}, allows us to estimate the \myred{average} OAM Eigenvalue $\ell_\pm=\pm4310$, which is within $1\sigma$ of the estimate achieved using the Sagnac effect. \myred{Two years ago in 2022 the first definitive observation of intrinsic longtiudinal OAM was reported \cite{Sarenac2022}.} We \myred{now} conclude that the Sagnac effect represents the first definitive detection of transverse neutron OAM, since it depends only on the projection of the OAM on the axis of rotation (equation \ref{SagnacPotential}), meaning that OAM must be present to explain a non-zero result. \par
Further, answering criticism raised in \cite{Cappelletti2018}, against the first experiments with neutron beam OAM \cite{Clark2015}, we propose to use OAM-rotation coupling to definitively detect longitudinal OAM. \myred{To accomplish this two technical difficulties need to be overcome (1) the sensitivity of the technique needs to be increased and (2) the rotation axis should be more closely alligned with the beam axis to measure the longitudinal OAM component.} The sensitivity of this method, using the earths rotation, is sufficient for detecting large quanta of OAM $|\ell|>10^3$, however it can be significantly improved by increasing the rotation frequency. A higher effective rotation frequency can be achieved by inserting a rotating Dove mirror (see for example \cite{Courtial1998,Leach2002,Geerits2021}) in the center of the instrument. \myred{Such devices are ubiquitous in photo-optics when it comes to measuring and sorting spin-orbit states \cite{Slussarenko2010,Wang2017}, hence we expect them to be equally useful in neutron optics, should they be implemented.} \myred{A pair of dove mirrors effectively rotate the image around the optical axis of the devices.} A low rotation frequency of 1 Hz, would increase sensitivity by $10^5$, compared to earths rotation. This method would increase $a_2$ and $c_{OAM}$, such that the systematic induced by the slight oscillations of the precession plane, becomes insignificant, \myred{since both parameters are proportional to the rotation frequency $\Omega$}. \myred{In addition the dove prism would address the second difficulty as well, since the axis of rotation for such a device can be chosen arbitrarily, including parallel to the propagation direction.} The Dove mirror could be made compact, albeit monochromatic, if mosaic crystals are used to produce reflections. \myred{Intuitively one may come to the conclusion that the dove mirror technique would only work if the cross section of the dove mirror is of a similar size as the neutron coherence length, owing to the fact that a neutron propagating off-axis but parallel to optical axis of the dove mirror, in its own frame of reference is not only rotated around its propagation axis, but also translated on a circle around the optical axis of the dove mirror. However intrinsic longitudinal OAM is translation invariant. In addition we have shown that the Sagnac method measures the OAM difference between two states which is translation invariant. As a result we conclude that the ratio between the cross section of the mirror and the neutron coherence length is irrelevant to the success of this technique.} \myred{Instead in SESANS interferometers it is important that the optical axis of the dove mirrors is precisely aligned with the optical axis of the interferometer. SESANS is designed to measure ultra small angle scattering, hence the alignment precision must exceed the instrument resolution. Luckily the resolution is proportional to the spin-echo length $q_{min}\propto\frac{1}{\delta_{SE}}$ and figure \ref{EvenOdd}, demonstrates that a low spin echo length produces an optimal superposition of $\ell=\pm 1$. Thus for a typical thermal neutron beam the alignment of the dove prism must be to within $0.1-1$ degree of the optical axis of the interferometer, so as to not introduce artifacts.} We postulate that dove mirrors will play an important role in neutron OAM optics for OAM manipulation and detection, since as opposed to the scattering methods reported on in \cite{Afanasev2019,Afanasev2021,Jach2022}, \myred{which depend on still unknown matrix elements}, the efficiency of a Sagnac based method is independent of other attributes of the neutron wavepacket such as \myred{transverse momentum and} coherence length. \myred{In addition the Sagnac method ought to provide a faster method than the interferometric methods used in for example \cite{Sarenac2022,Geerits2023}, since these approaches require detectors with spatial resolution to resolve the OAM phase structure. Due to the low flux of neutron beams this requirement of spatial or angular resolution results in a larger integration time compared to the Sagnac method.} \myred{Since the matrix elements, which determine the scattering amplitudes of twisted neutrons from nuclei are still unknown, the Sagnac method may be used to calibrate twisted scattering techniques.} \myred{(Dove) mirrors may also be used to increase the sensitivity of our measurement to extrinsic transverse OAM. A mirror at the center of our instrument rotating around the vertical of the instrument effectively acts as a rotation of the instrument. As a result one can simulate faster rotating frequencies than that provided by the earth.}\par 
As pointed out in previous sections our technique uses spin-orbit entanglement to imprint OAM dependent phases on the spin (see for example eq. \ref{Entangled}) and to characterize these phases by measuring the spin projection. This is somewhat analogous to the entangled optical interferometer reported on in \cite{Silvestri2024}, which makes use of inter-particle entanglement, while we report on intra-particle or mode entanglement. The $1\sigma$ precision achieved in our experiment corresponds to a rotational sensitivity of $5.1 \mu rad \cdot s^{-1}$, similar to what is reported in \cite{Silvestri2024}, which to our our knowledge is the most sensitive measurement of the Sagnac effect using entanglement.
\section{Acknowledgments}
N.~G. and S.~S. acknowledge funding from the Austrian science fund (FWF), Project No. P34239, in addition N.~G. and W.~M.~S. are supported by the US Department of Energy (DOE) grant DE-SC0023695. W.~M.~S. and K.~S. acknowledge support from the US National Science Foundation (NSF) grant PHY-2209481 and the Indiana University Center for Spacetime Symmetries. The authors extend their gratitude to G{\o}ran Nilsen for supplying the ramped $\pi/2$ flipper. Experiments at the ISIS Neutron and Muon Source were supported by a beamtime allocation RB2410116 \cite{LarmorData} from the Science and Technology Facilities Council.

\section{References}
\bibliographystyle{unsrt}
\bibliography{SpinOrbitBibliography}
\end{document}